# Precision determination of corotation radii in galaxy disks: Tremaine-Weinberg v. Font-Beckman for NGC 3433

John E. Beckman[1,2,3], Joan Font[1,2], Alejandro Borlaff[1,2], Begoña García-Lorenzo[1,2]

[1] Instituto de Astrofísica de Canarias, c/Vía Láctea, s/n, E38205, La Laguna, Tenerife, Spain; jeb@iac.es, jfont@iac.es, asborlaff@iac.es, bgarcia@iac.es
[2] Departamento de Astrofísica, Universidad de La Laguna, Tenerife, Spain.
[3] Consejo Superior de Investigaciones Científicas., Spain

Abstract

Density waves in galaxy disks have been proposed over the years, in a variety of specific models, to explain spiral arm structure and its relation to the mass distribution, notably in barred galaxies. An important parameter in dynamical density wave theories is the corotation radius, the galactocentric distance at which the stars and gas rotate at the same speed as the quasi-static propagating density wave. Determining corotation, and the pattern speed of a bar have become relevant to tests of cosmologically based theories of galaxy evolution involving the dynamical braking of bars by interaction with dark matter haloes. Here comparing two methods, one of which measures the pattern speed and the other the radius of corotation, using two instruments (an integral field spectrometer and a Fabry-Perot interferometer) and using both the stellar and interstellar velocity fields, we have determined the bar corotation radius, and three further radii of corotation for the SAB(s)b galaxy NGC3433. The results of both methods, with both instruments, and with both disk components give excellent agreement. This strengthens our confidence in the value of the two methods, and offers good perspectives for quantitative tests of different theoretical models.

1. Introduction.

Determination of the corotation radius in a galaxy disk, a key component of density wave theory (Lindblad 1938, Lin & Shu 1964) has been relevant to cosmology since the prediction by Weinberg (1985), Hernquist & Weinberg (1992), and Tremaine & Ostriker (1999) that the bars in barred galaxies should have been dynamically braked by their dark matter halos on timescales significantly less than the Hubble time. This prediction was focused by Debattista & Sellwood (2000) and by Sellwood & Debattista (2006) into a quantitative prediction of this effect, on the ratio of the corotation radius to the length of the bar. The difficulty of determining corotation radii has been gradually overcome during the last decade. A known method, first proposed by Tremaine & Weinberg (1984), hereinafter TW, and first used on the S0 galaxy NGC 936 (Merrifield & Kuijken, 1995) combines the surface density distribution of stellar light from photometric imaging with line of sight velocity distribution using long slit spectra. The technique has been usefully applied to a couple of tens of objects, notably by Aguerri et al. (2003), Corsini et al. (2007), Aguerri et al. (2015), with uncertainties which lie in the range 10% to 40%. An interesting generalization of the method to galaxies with multiple pattern speeds and corotation radii was presented in Meidt et al. (2008). The Tremaine-Weinberg method has been considered standard, but has been limited in applicability due to long observation times with conventional long-slit spectra. Integrated field spectra have recently improved the situation, but over relatively restricted fields. For example, the PPAK IFU with which the CALIFA survey was conducted (Sanchez et al. 2012) has "an extremely wide field" of only 1.3 arcmin$^2$. Debattista and Williams (2004) showed that a Fabry-Perot spectrometer could be used for 2D absorption line spectroscopy of a stellar population, obtaining very precise TW results for NGC 7079, but their field was 1.4 arcmin$^2$ compared to the field of the GHαFaS Fabry-Perot spectrometer used in the present work, of 12 arcmin$^2$.

More recently two of the present authors presented the Font-Beckman method, (Font et al. 2011, 2014a, 2014b), hereinafter FB, based on the change in phase, at corotation, of the radial component of the velocity in the disk, from inwards to outwards, or vice versa first brought out explicitly by Kalnajs (1978). This was used to determine corotation radii in over 100 galaxies, with uncertainties typically in the range 5%-10%. It was applied to 2D velocity fields obtained in Hα emission at good spatial (~1 arcsec) and spectral (~ 8 km s$^{-1}$) resolution, i.e. using the ionized gas as a tracer. Although FB is considerably faster, and can in practice reach larger disk radii, than TW, it is most useful for star-forming, i.e. later type galaxies where there is sufficiently extended Hα emission and is complementary to TW. Indeed TW is intended for use with the stellar components of galaxies, as it ideally requires continuity, while FB is designed for use with the ISM, and does not entail this. The aim of the work presented here was to find a galaxy which could serve as a test-bench, where the data allowed us to apply both techniques, so that we could compare them directly, helping observers to judge where best each could be applied. We also hoped that using both methods on a given object we might optimize the precision with which corotation could be measured, again giving hints for wider applications. In section 2 we describe the observations, in section 3 we give a brief summary of the two methods being compared, in section 4 we show the results of the application of both methods, and in section 5 we discuss the results and draw conclusions.

2.Observations.

NGC 3433, an SAB(s)b galaxy (Buta et al. 2015) was chosen because its declination, at ~+10°, makes it observable from the latitude of the VLT at Paranal, and also from the William Herschel Telescope (WHT) on La Palma. This means that we could use and compare 2D spectral data from MUSE (Bacon et al., 2014) on the VLT and from our own instrument GHαFaS (Hernandez et al. 2008) on the 4.2m WHT. A further criterion is that the full optical disk of NGC 3433 fits a single GHαFaS field of 3.4 x 3.4 arcmin. For comparison in Figure 1 we show the field of MUSE, (red square) compared with the field of GHαFaS (yellow square). We can see one advantage of GHαFaS: its field, over ten times the area of the MUSE field. The internal field of

NGC3433, including the bar region, is however covered by the field of MUSE. So, the GHαFaS field is needed for the interstellar emission in the outer spiral arms.

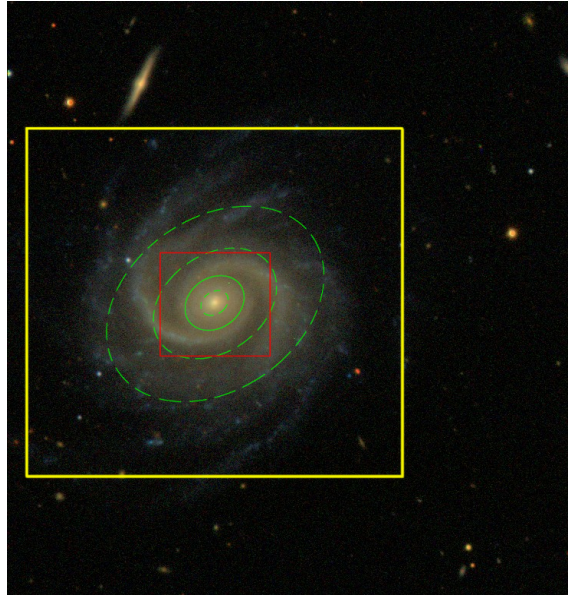

Figure 1. NGC 3433 (image from SDSS). Superposed: MUSE field (red square), GHαFaS field (yellow square); Projected corotation circle for the bar (solid green line); the other measured corotation radii (dashed green lines).

In Table 1 we give the basic parameters of NGC 3433. These include the position and the morphological type from the standard literature, and also the morphological parameters we have inferred during the course of the work reported in this article. These include the position angle, the inclination angle, and the length of the bar. Our kinematic analysis, explained below, carried out using a sequence of concentric tilted rings, was the basis of most of the values in Table 1, as shown, but we used the Spitzer 3.6 μm image to measure the bar length (this was also used by Herrera-Endoqui et al. 2015), and also to derive a purely morphological estimate of the disc position angle. We note here that the measurement of the inclination angle from three distinct data sets yields very consistent values. The most precise of these, using the full disc Hα kinematics from GHαFaS, has an intrinsic error of only $0.8^o$ and lies well within the error limits of the two values obtained from the inner disc from the MUSE data. The position angle shows a significant variation of some 9 degrees between the MUSE values and the GHαFaS value. We attribute this to the presence of the bar which must affect both the isophotes and the isovels in the inner part of the galaxy, the part observed with MUSE. This inference is supported by the measured angular separation between the bar and the disc major axes, which is close to $10^o$. In our TW derivation we have used the inner position angle for the MUSE data and the GHαFaS inner disc data, and the outer position angle for the GHαFaS outer disc data. We should make it clear that this discrepancy does not represent a variational uncertainty in the determinations Where we show these determinations, below, we will deal further with the measurement uncertainties.

Table 1. Parameters of NGC 3433

| Parameter | Value | Reference |
|---|---|---|
| Morphological type | SAB(s)b | Buta et al. 2015 |
| RA (J2000) | 10h52m03.870s | |
| DEC (J2000) | +10d08m53.92s | |
| Inclination Angle | 34.2 ± 1.4º [1] | This work |
| Position Angle | 24.8 ± 1.1º [1] | This work |
| | 23.5 ± 3.7º [2] | This work |
| Bar Length | 15.5 arcsec | Herrera-Endoqui et al. 2015 |
| | 15.4 ± 1.2 arcsec [1] | This work |

*Notes* (1) These are the average values obtained fitting tilted ring models to the three velocity maps obtained from each of the velocity fields shown in Figure.2. (2) This value is obtained by ellipse fitting to the 3.6 μm Spitzer model.

2.1 Observations with GHαFaS

GHαFaS is a Fabry-Perot interferometer at the Nasmyth focus of the WHT. It has a free spectral range of 8.5 A, corresponding to 390 km·s$^{-1}$ at the rest wavelength of Hα, divided into 60 independent channels yielding a formal spectral resolution of 6.5 km·s$^{-1}$. The observations were performed on March 27$^{th}$ 2015, with an average value for the seeing of 1.2 arcsec. The angular resolution was seeing limited, with a pixel scale of ~0.2 arcsec/pixel. The observing time on NGC 3433 was 3 hours, the reduced observations (see Hernandez et al. 2008, Font et al. 2011) yield a wavelength calibrated data cube. We apply Principal Component Analysis (PCA, Steiner et al. 2009) to this cube. PCA conserves flux, and yields a cleaned cube, in which the spectra are smoothed and the noise minimized, making it easier to measure the parameters of the emission line spectra (line center, FWHM and integrated flux) derived from the cube. After applying PCA a 2D Gaussian smoothing, with a FWHM of 5 pixels, was applied, and the moment maps (corresponding to integrated surface brightness, peak velocity, and velocity dispersion) were derived. We need only the peak velocity map for our purpose here, which is displayed in panel a of Figure 2.

2.2 The MUSE observations.

We used archive observations from the Multi Unit Spectrograph Explorer (MUSE), on the Very Large Telescope (VLT) of the European Southern Observatory (ESO). The wavelength coverage is from 4650 to 9300 Å over a field of view of 1x1 arcmin$^2$. NGC 3433 was observed with MUSE on 20$^{th}$ February 2014 (Program ID: A-9100), using the wide field mode, with an exposure time of 3900 seconds divided among 13 exposures. The atmospheric seeing, which limited the angular resolution, ranged from 0.6 to 0.8 arcsec. The data were published on July 28$^{th}$ 2014 and were reduced to a single data cube using MUSE data reduction software (pipeline version v1.0.1) in the Reflex environment. This combines the calibration files to correct the science frames for bias, dark, flat, geometric calibration and sky emission, while performing the wavelength and photometric calibrations. The final data product is a cube of 315 x 316 x 3682 spaxels, with effective angular size 0.2 x 0.2 arcsec and a formal spectral resolution of 1.25 Å corresponding to some 50 km·s$^{-1}$. With these data, as with the GHαFaS data described above, significantly higher effective spectral resolution can be achieved given a sharp enough line or set of lines, and a high enough signal to noise ratio. Note that the PCA technique is also applied to the MUSE data.

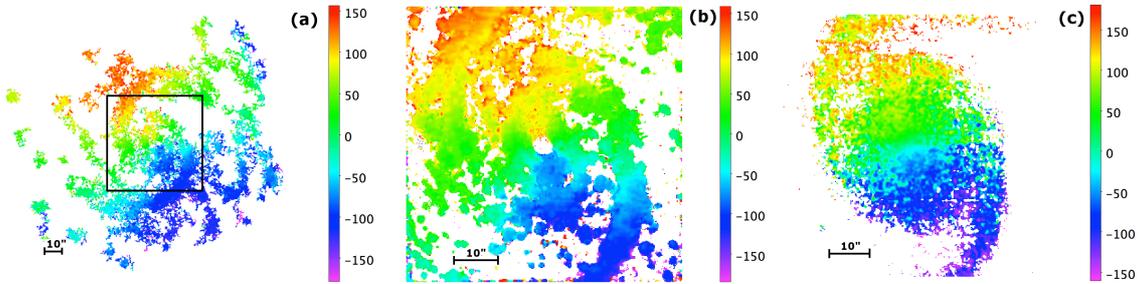

Figure 2. Panel a: velocity map of NGC 3433 using 1$^{st}$ moment map of Hα emission in the FP data cube from GHαFaS. The box in black shows the size of the MUSE data. Panel b: velocity map of Hα emission from the central armin$^2$ from the MUSE data cube. Panel c: velocity map of the stellar component from the MUSE data.

As an initial step in the kinematic analysis we need to derive the rotation curve of the galaxy from the velocity map, using the ROTCUR task of the GIPSY[1] package. This technique yields a circular velocity for each of the tilted rings employed in the method, and these points can then be plotted against the galactocentric radii of the rings. We used the method on each of the velocity maps shown in Figure 1, for stars and gas in the MUSE data and for gas alone in the GHαFaS data. We first plotted separate rotation curves for each of the three data sets using the best fit polynomials with the expression:

$$V_{rot}(r) = a \cdot \frac{r/b}{\left[1/3 + 2/3 \left(r/b\right)^{1/2}\right]^3}$$

where a is a velocity with a value which turns out to be close to the maximum velocity of the curve, and b is a radius whose value is close to that at which this velocity is attained. As a second step we plotted all the points derived from the three velocity fields in a single graph, and performed a single polynomial fit to the combined data set. This is shown in Fig. 3, together with its associated band of uncertainty. We used the curve in Fig. 3 as a starting point for the FB derivation of the resonance radii as described below.

---

[1] https://www.astro.rug.nl/~gipsy/

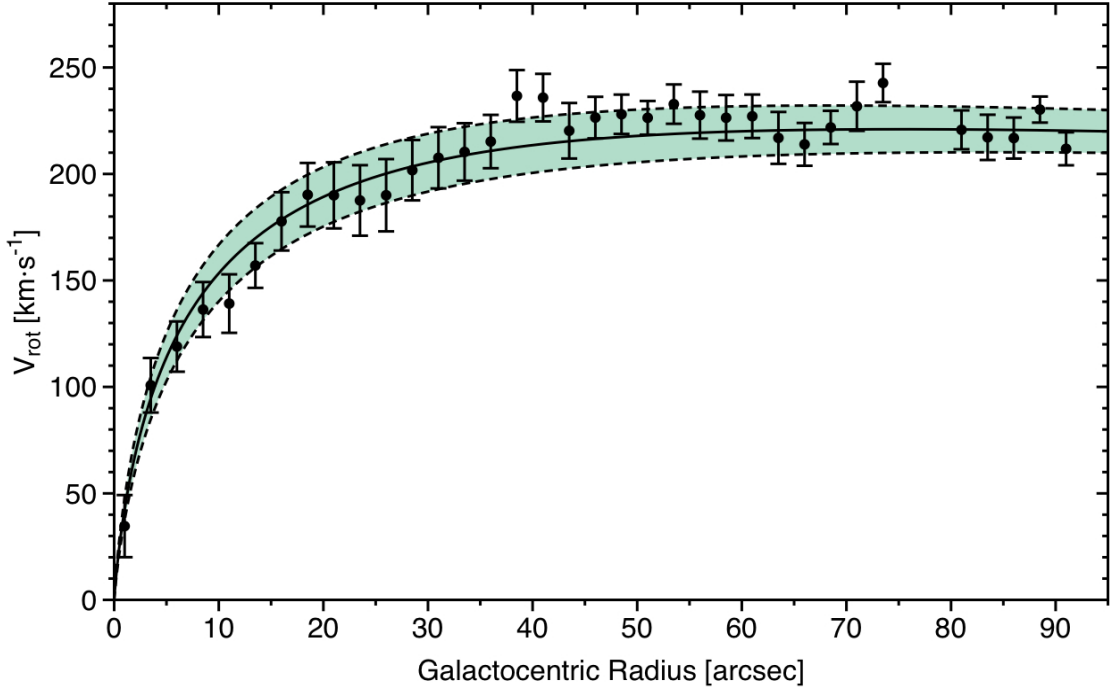

Fig. 3. The rotation curve for NGC 3433. The points are those derived from the individual tilted rings used in the analysis of the velocity fields shown in Fig. 2; we have combined the data from the three fields. The smoothed rotation curve used to derive the resonance radii using FB, obtained using a polynomial fit, is shown as the continuous curve, and the range of uncertainty is shown as the shaded area around this curve.

3. The two methods to be compared

The two methods we used to derive the corotation radii for NGC3433 have been described in detail in the literature, so here we will give a very brief summary of each. An important point, is that the methods measure two different parameters. The TW method measures the pattern speed, (in the original formulation, of the bar in a barred galaxy), from which the corotation radius can be derived using the rotation curve. The FB method determines a corotation radius directly, and its corresponding pattern speed can then be determined, again from the rotation curve. Finding agreement between the two measured values would strengthen the validity of both methods. To find the pattern speed using TW (see Merrifield & Kuijken 1995) we need to measure two quantities: the line of sight velocity of the stellar population, measured along a set of long spectral slits parallel to the major axis of the disc, and the continuum emission from that population measured over the same field, using the expression:

$$\Omega_P = \frac{1}{\sin(i)} \cdot \int \frac{l(x) \cdot [v_{los}(x) - v_0]}{l(x) \cdot [x - x_0]} dx$$

where the integrals are performed along the slit, $v_{los}(x)$ is he mean line of sight velocity of the stars at position x on the slit, $v_o$ is the systemic velocity of the galaxy, the square brackets indicate the average value along the slit weighted by $l(x)$ the luminosity of the stars at each point, and $i$ is the angle of inclination of the galaxy disk to the line of sight. To improve the signal to noise ratio the process can be repeated along as many slits as can be observed, and an effective mean derived. In principle this method assumes a continuity equation for the population whose velocity is measured, which would imply that only an older stellar population should be used. We will see that this condition may be relaxed and TW can be applied to the velocity field of the interstellar gas. However it has mainly been used for earlier types of disk galaxies because dust obscuration must affect the photometric continuity of the stellar continuum signal.

To find a corotation radius using FB we use the change in phase of the non-circular (radial) velocity vector across corotation. Using a 2D map of the line of sight velocity, we first derive a mean rotation curve, and use this to form a 2D circular velocity model. This model is subtracted off from the initial field leaving a residual field of (projected) non-circular velocity. The third step is to identify the positions in the map where the residual velocity is zero and, (to avoid problems of noise and projection effects), to eliminate the sub-set of these pixels where the change in radial velocity, from inflow to outflow or vice-versa, has an amplitude smaller than twice the rms uncertainty in the velocity measurements. The fourth step is to plot a histogram of the numbers of the remaining pixels against galactocentric radius. This histogram shows very well defined peaks, which we identify as corotation radii.

From our previous studies using FB (Font et al. 2011, 2014a 2014b) we have shown that there is usually more than one corotation in a galaxy, and we have always been able, using the morphology, to assign each corotation to a feature: a main bar, a nuclear bar, or an annular section of the spiral arms. This has produced results akin to those of Meidt et al. (2008) who modified the simplest form of TW to assign several corotation radii to identified morphological features of the galaxies over concentric annuli of galactocentric radius. Because FB normally uses the emission from the interstellar medium it is most suited to later type galaxies, and is complementary to TW. However for the galaxy under study we could use both methods, giving us the possibility to compare them.

4. Measurements of corotation radii.

4.1 Using TW

The MUSE spectra give a wide range of stellar absorption lines, but also interstellar Hα emission, so we could use the 2D spectral map to derive pattern speeds from both types of velocity fields, and then go on to infer the corotation radii. Without trying to segment the data radially TW gave us a clear pattern speed for the main bar (which is quite oval) using both stars and gas, with the inferred corotation radius at 18.3±4.1 arcsec using the Ca near-IR triplet in absorption for the stellar component,

(following the technique of using multiple lines to enhance the effective velocity precision for TW described in García-Lorenzo et al. 2015), and at 18.2±3.5 arcsec using interstellar Hα emission. It was easier to make radial segments with the GHαFaS observations, and we could measure a wider set of corotations using TW and GHαFaS. These yielded a corotation at 17.9±4.5 arcsec, when the GHαFaS data are confined to the same region as the MUSE observations, which is in good agreement with the MUSE value using TW, and a corotation at 58.9±8.2 arcsec using the complete GHαFaS map.

In Fig. 4 we show the TW plots using both instruments. In the left panel we show detailed results for the MUSE and GHαFaS observations. Each point represents the value for a pseudo-slit of the ratio of the integrated value of the line of sight velocity and the integrated distance, along the slit, both weighted by the luminosity. The pattern speed is derived from the slope of the graph. The left panel shows plots using stars and gas from MUSE, and the central zone, which is dominated by the bar, from the GHαFaS data. The right panel shows the full disc fit for the GHαFaS data. Luminosity weighting for the Hα observations used the nearby stellar continuum.

We note here that the quoted uncertainties in the values of the pattern speeds obtained using TW combine the measured uncertainty in the inclination angle of the disk, the uncertainty in the position angle of the inner zone for the TW plots in Fig. 4, left panel, the uncertainty in the position angle of the outer disk for the TW plots in Fig 4, right panel, and the uncertainties in the slopes of each of the TW plots presented in both panels of Fig. 4. When converting a pattern speed to the corresponding value of the corotation radius the uncertainty in the frequency curve was also taken into account, but, as will be seen in Figure 5, the resulting estimated uncertainties are clearly conservative.

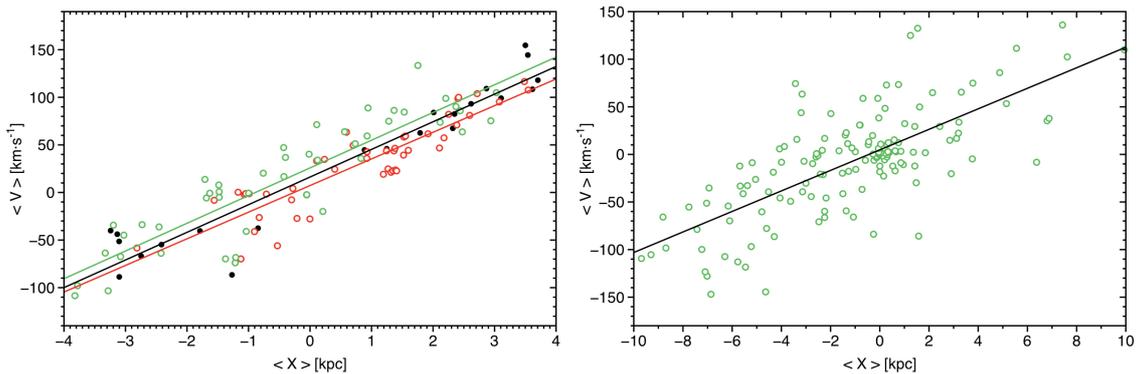

Figure 4. TW plots of the luminosity weighted integrals of the line of sight velocity along selected pseudoslits parallel to the major axis of NGC 3433 against the luminosity weighted distance along the slit. Each point represents a value for a single slit. The zero point for a given slit is the point closest to the rotational center of the galaxy. The left panel includes the stellar spectra from MUSE (black circles), Hα emission from MUSE (open circles in red), and the center zone with data from GHαFaS (open circles in green); lines in red, black and green correspond to the linear fit for the stellar and

Hα data from MUSE and the ionized gas from GHαFaS, respectively. The right panel shows the plot for the full disk with data from GHαFaS.

4.2 Using FB

Using the MUSE map to derive corotation radii directly with FB, and first taking the interstellar Hα emission line map, we found three values: 8.9±0.5 arcsec, 18.5±1.4 arcsec, which is in good agreement with the TW value above, and 31.5±3.0 arcsec. If we use FB with the stellar absorption spectral map from MUSE, we find only one corotation, at 18.3±2.7 arcsec, in excellent agreement with the three values obtained with TW, above. Turning to the GHαFaS map and deriving corotations with FB we find one at 8.2±0.6 arcsec, another at 17.8±1.9 arcsec, a third at 31.9±2.4 arcsec, and a fourth at 59.5±2.5 arcsec. The radius of the innermost corotation is in excellent agreement with the values found using FB with MUSE. The second corotation radius is in excellent agreement with the values found using TW with both gas and stars, from the MUSE map, as well as with the values found using FB with both stars and gas in MUSE. The third corotation radius in in excellent agreement with the value found using FB with gas, from the MUSE data. The fourth corotation radius derived using FB and gas, with GHαFaS cannot be compared with any other value, because this radius is well outside the MUSE field, and GHαFaS does not obtain a stellar velocity map. All the observed values of the 4 resonances are shown in Table 2, where we can see the agreement between the two methods in all cases, well within the uncertainty limits, and also the agreement between the values obtained using the kinematics from the stellar spectra and those using the kinematics from the interstellar spectra.

**Table 2.** Corotation radii and pattern speeds for NGC 3433

| Resonance | Method | Corotation Radius arcsec | kpc | Pattern Speed km s$^{-1}$ kpc$^{-1}$ |
|---|---|---|---|---|
| 1 | $FB_g(\gamma)$ | **8.2±0.6** | 1.3±0.1 | 85.1±3.9 |
| 1 | $FB_g(\mu)$ | **8.9±0.5** | 1.4±0.1 | 81.0±3.8 |
| 2 | $FB_g(\gamma)$ | **17.8±1.9** | 3.6±0.4 | 49.2±3.7 |
| 2 | $TW_g(\gamma)$ | 19.1±2.1 | 3.9±0.5 | **46.7±3.7** |
| 2 | $FB_g(\mu)$ | **18.5±1.4** | 3.7±0.3 | 47.8±3.0 |
| 2 | $TW_g(\mu)$ | 20.0±2.7 | 4.0±0.6 | **45.0±4.2** |
| 2 | $FB_s(\mu)$ | **18.3±2.7** | 3.7±0.5 | 48.2±5.2 |
| 2 | $TW_s(\mu)$ | 18.1±3.0 | 3.7±0.6 | **46.7±4.9** |
| 3 | $FB_g(\gamma)$ | **31.9±2.4** | 6.4±0.5 | 30.9±2.0 |
| 3 | $FB_g(\mu)$ | **31.5±3.0** | 6.4±0.6 | 31.3±2.6 |
| 4 | $FB_g(\gamma)$ | **59.5±2.4** | 12.1±0.5 | 17.6±0.7 |
| 4 | $TW_g(\gamma)$ | 60.6±7.2 | 12.3±1.4 | **17.3±1.8** |

*Notes.* Second column gives the method used: $TW_s$ refers to Tremaine-Weinberg method using the stellar component; $TW_g$, Tremaine-Weinberg method using the ionized IS gas; $FB_g$, Font-Beckman Method using the ionized IS gas; $FB_s$, Font-Beckman method using the stellar component. The instrument used is indicated in brackets, γ means GHαFaS and μ means MUSE. Numbers in boldface are the primary

parameters derived by the method used (pattern speed for Tremaine-Weinberg, corotation radius for Font-Beckman).

It is instructive to show how the values of those corotations where we could use both TW and FB compare, taking into account the uncertainties in their derivations. In each panel of Figure 5 we have plotted the frequency curve ($\Omega$ against galactocentric radius) obtained using the relevant fitted rotation curve from Fig. 3; the uncertainties associated with this curve are shown as shaded regions in green, which are calculated taking into account the uncertainties of the rotation curve and the standard deviation of the fitting parameters to the rotation curve. For a given corotation we have plotted the radius, derived directly using FB, as a vertical dashed line, with the uncertainty as a blue shaded surrounding bar. For the comparison we have plotted the pattern speed derived using TW as a horizontal dashed line, with the associated uncertainty as a pink shaded surrounding bar. In the three panels we have shown these measurements for, respectively the GH$\alpha$FaS H$\alpha$ field, the MUSE H$\alpha$ field, and the MUSE stellar field. We can see that for all the corotations the crossing points of the vertical bars with the $\Omega$ curves are in agreement with the crossing points of the horizontal bars with the curves, and the two crossing points are in fact indistinguishable in all the four cases shown. This implies that the uncertainties which we have computed as described above are in fact conservative. It also implies that the two methods give identical results, which we take as strong support for FB given that TW is an established technique.

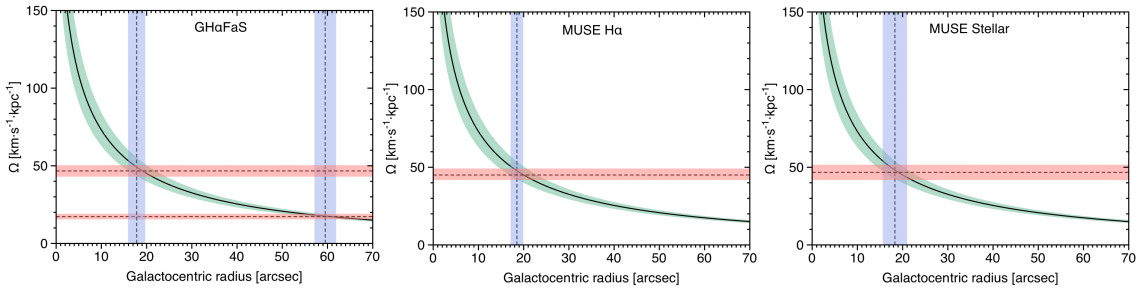

Figure 5. Left panel. The frequency curve (solid line) with its uncertainty shaded in green is plotted. The two shaded vertical regions (in blue) indicate the corotation radii (vertical dashed line) including their uncertainties, directly calculated with the FB method, corresponding to resonances (2) and (4) of Table 2, while the pattern speed values directly calculated with TW method are shown as horizontal shaded regions in red. Central panel. The same as the left panel but for the resonance found in the MUSE velocity map of the H$\alpha$ emission (panel b of Figure 2). Right panel. The same type of plot is shown here for the MUSE stellar velocity map.

In Figure 6 we have summarized the information in Table 2 for ease of visualization. In the left panel we have shown the corotation radii against the method used to obtain each value. We show the results of the FB determinations in red, as these are the direct measurements, and the results of the TW determinations in black, as these are derived by combining the measured pattern speed with the angular velocity of the rotation curve. In the right panel we have shown the pattern speeds. In this case the direct determinations using TW are plotted in red and the values using FB, which are obtained by combining the measured pattern speeds with the angular velocity of the rotation curve are plotted in black. It should be clear that the larger corotation radii correspond to the smaller values of pattern speed, so the plots are in this sense inversely related.

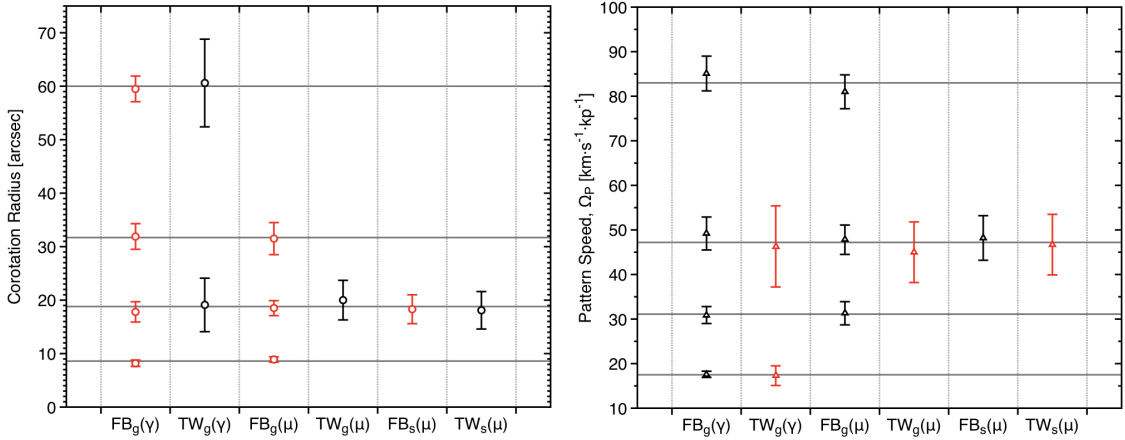

Fig 6 Left-hand panel. Corotation radii either measured directly using FB (red circles) or inferred from TW (black circles). The horizontal line (in grey) marks the averaged value of the corotation radius for each resonance. Notation: γ means a GHαFaS determination, μ means a Muse determination; subscript 's' means an interstellar gas measurement, subscript 's' means a stellar measurement. For the corotation associated with the bar there are six separate determinations. Right-hand panel. The equivalent information to that presented in the left panel but now shown in terms of pattern speed. In this diagram the direct measurements are from TW (red triangle) and the indirect determinations from FB (black triangle). The remaining notation is the same as in the left panel.

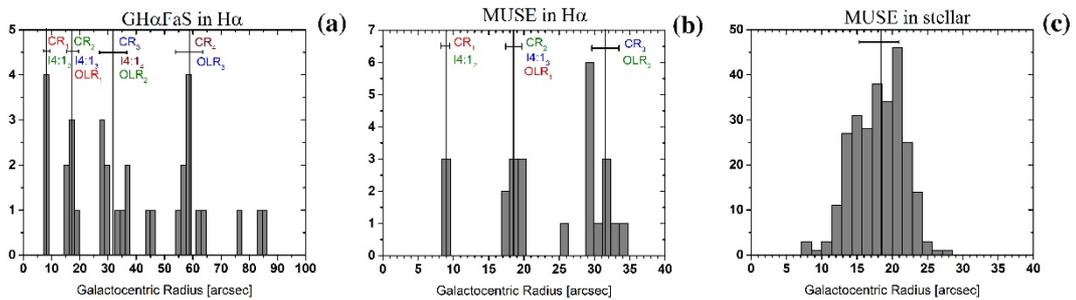

Figure 7. FB histograms with peaks corresponding to the measured corotation radii of NGC 3433: (a) using the GHαFaS velocity field in Hα. Each corotation is numbered and coded with different color. The radius of the corresponding Outer Lindblad resonace (OLR) and the ultraharmonic inner 4:1 resonance (I4:1) are also plotted and appear colored and numbered according to its associated corotation, in order to show the coupling between interlocked resonances as described in the text. (b) the same as panel (a) but using the MUSE velocity field in Hα. (c) using the MUSE velocity field from the stellar absorption spectrum. The solid vertical lines show the radial position of the resonance and the horizontal segments on the top indicate the uncertainty.

Fig. 7 shows the histograms which we used to derive the corotation radii, using data from both instruments and the velocity fields of both gas and stars. On the left is the histogram from the GH$\alpha$FaS velocity field. We detect four clear peaks, whose radii are defined by the Gaussian fits to the histogram. In the center panel we have the equivalent histogram from the MUSE H$\alpha$ emission velocity field. The resonance peaks agree very well with those from GH$\alpha$FaS although given the restricted angular size of the MUSE field we find only the inner three resonances. On the right we show the MUSE histogram produced using the absorption lines of the stellar population. We can see that there is only one peak, at the corotation corresponding to the bar, which we would expect to be the strongest resonant feature. This peak is much wider than those in the previous figure, and we attribute this to the fact that the stellar orbits are "more three-dimensional" than the gas orbits. The stars have a greater scale height above the disk plane than the gas, and a significant component of velocity perpendicular to the plane. Our technique of finding zeros in the radial velocity is thus partly compromised by false zeros due to components perpendicular to the plane. This has not, however, affected the value of the corotation radius obtained using FB with the stellar component, determined by the peak of the distribution, as listed in Table 2.

In Figure 7, we also show how the resonances are coupled in the sense we reported in Font et al. 2014a. This means that given two pattern speeds, for example $\Omega_1$ = 85.5±4.1 km s$^{-1}$ kpc$^{-1}$ and $\Omega_2$ = 46.8±3.3 km s$^{-1}$ kpc$^{-1}$, with corotations at 8.9±0.5 arcsec and 18.5±1.4 arcsec, respectively (see Table 2), it turns out that the radius of the Outer Lindblad Resonance of $\Omega_1$ (OLR$_1$) is consistent with the corotation radius of $\Omega_2$ (CR$_2$), within uncertainties, and also the Inner 4:1 resonance of $\Omega_2$ (I4:1$_2$) occurs at the same radial position as the corotation of $\Omega_1$ (CR$_1$). This interlocking pattern is displayed in Figure 7, panel b, in which the radial positions of each resonance is colored and numbered according to its associated corotation. This figure also shows that we have identified two interlocking pattern between the three resonances found: between the first and the second resonance, and between the second and the third resonance; finally using the outermost resonance found only with GH$\alpha$FaS in H$\alpha$ we also detect the same characteristic coupling pattern between the third resonance and the fourth, outermost resonance.

Conclusions

Even though density wave theory has deep historical roots (Lindblad 1938, Lin & Shu 1964) and innumerable subsequent research contributions, there are several aspects of the theory which could still benefit from rigorous observational support. Foyle et al. (2011), for example, put question marks over the long-term stability of spiral arms, based on measured sequences of star formation indicators across the arms, but the predictions tested were for density wave models with a single corotation radius, while Font et al. (2014) showed that virtually all the star-forming disk galaxies they observed showed multiple corotations. In this context, as well as to underpin attempts to use the rotation of bars in tests of cosmological models, it is valuable to quantify the parameters of a density wave system by comparing two independent methods for determining its basic parameters. We have done this here and give the following conclusions:

1. It is possible to determine pattern speeds and corotation radii of spiral galaxies quite accurately. The values of the formal uncertainty for the corotation radii obtained for NGC 3433 are of order 3%.

2. (a) We used two very different instrumental and telescope set-ups, with results which are the same for both.

(b) We used two methods which employ different observational input parameters and derive different output parameters. TW yields values for the pattern speed and FB yields values for the corotation radii. Using the rotation curve it is then easy to compute the other output parameter for each method. The results are in complete agreement.

(c) We used the velocity fields of the stellar component and of the ionized gas component for both TW and FB, with results which give full agreement.

When using the ratio of the corotation radius to bar length to test the hypothesis that bars in galaxies should be braked by dynamical friction Font et al. (2016) showed that the uncertainty in the measured length of the bar, rather than in the radius of corotation, is the limiting factor in deriving this ratio. Here we have gone further, showing that corotation radii in disk galaxies can be among the most precisely measurable parameters in disks, and that the general principles of density wave theory are strongly supported by these measurements.

Acknowledgements

We thank the anonymous referee for a series of perceptive comments which have caused us to make significant improvements in the article. The research was carried out with funding from project P/308603 of the Instituto de Astrofísica de Canarias. The William Herschel Telescope is operated on the island of La Palma by the Isaac Newton Group in the Spanish Observatorio del Roque de los Muchachos of the Instituto de Astrofísica de Canarias. Based on observations collected at the European Organisation for Astronomical Research in the Southern Hemisphere under ESO programme A-9100. B.G-L acknowledges support from the Spanish Ministerio de Economia y Competitividad (MINECO) by the grant AYA2015-68217-P.